# First step toward multi machine ELM energy scalings and extrapolations to SPARC and ITER

R. Perillo[1]*, A. Redl[2]*, T. Eich[2], C.J. Lasnier[3], A. Nelson[4], R. Rizkallah[1], D. Silvagni[5], A. Stagni[6], J.A. Boedo[1], A. McLean[3], P. Traverso[7], N. Vianello[6], the DIII-D team[a], JET contributors[b], the ASDEX Upgrade team[c], the TCV team[d] and the EUROfusion Tokamak Exploitation team[e]

[1]University of California San Diego, La Jolla, CA 92130, USA
[2]Commonwealth Fusion Systems, Devens, MA, USA
[3]Lawrence Livermore National Laboratory, Livermore, CA 94550, USA
[4]Columbia University, New York, New York 10034, USA
[5]Max–Planck-Institut für Plasmaphysik, 85748, Garching, Germany
[6]Consorzio RFX, (CNR, ENEA, INFN, Università di Padova, Acciaierie Venete SpA), Corso Stati Uniti 4, 35127 Padova, Italy
[7]Oak Ridge Associated Universities, Oak Ridge, TN, USA
[a]See author list of C. Holcomb et al 2024 *Nucl. Fusion* **64** 112003
[b]See author list of C. Maggi et al 2024 *Nucl. Fusion* **64** 112012
[c]See author list of T. Pütterich et al 2026 *Nucl. Fusion* **66** 116002
[d]See author list of C. Theiler et al 2026 *Nucl. Fusion* **66** 112023
[e]See author list of N. Vianello et al 2026, *Nucl. Fusion* **66** 116010
*These authors contributed equally to this work.

*Abstract*

It is shown that the ELM energy loss normalized by the plasma stored energy ($\Delta E_{ELM}/W_{plasma}$) for high-density small/QCE ELM regimes scales inversely with the separatrix turbulence parameter $\alpha_t$. In contrast, the neoclassical electron collisionality at the pedestal top, $\nu^*_{e,neo}$, expected to regulate $\Delta E_{ELM}/W_{plasma}$ according to the Loarte scaling (*Plasma Phys. Control. Fusion* 2003 45 1549), does not adequately capture $\Delta E_{ELM}/W_{plasma}$ data for peeling-ballooning-limited type-I ELMs and ballooning-limited small/QCE ELMs, limiting its applicability for extrapolation to one scenario window. A multi-machine database including seven tokamaks and with $\Delta E_{ELM}/W_{plasma}$ ranging from 0.5% to 14%, has been analyzed. A regression analysis on only type-I ELMs yields $\left(\frac{\Delta E_{ELM}}{W_{plasma}}\right)_{Type-I}[\%] = 6.8 * T_{e,ped}^{0.03} n_{e,ped}^{-0.4} \kappa^{-0.4} R_{major}^{0.4}$, corresponding to $\Delta E_{ELM}/W_{plasma}$ =4.5% for nominal SPARC pedestal parameters and 12% for the ITER D-T Q=10 scenario. For the small/QCE ELM class, however, as $\alpha_t$ increases, the pedestal moves toward a ballooning-limited boundary, the toroidal mode number increases, the ELM frequency rises following the scaling $f_{ELM} = 46e^{(2.25*\alpha_t)}$, and $\Delta E_{ELM}/W_{plasma}$ decreases via the relation $\frac{\Delta E_{ELM}}{W_{plasma}}[\%] = 1.6e^{-(\frac{\alpha_t}{2})}$. For SPARC QCE-relevant $\alpha_t$=0.86 and ITER high-fueling scenario $\alpha_t$= 0.64, the scaling favorably predicts $\Delta E_{ELM}/W_{plasma}$ of 1.0% and 1.2%, respectively, with values below 1% if the small/QCE ELM regime is pushed beyond $\alpha_t$ >1. The small/QCE ELM-fitted results represent an initial step toward future analysis on broader datasets, which will be necessary to improve the accuracy of projections for future reactor-relevant scenarios.

*1. Introduction*

Edge localized modes (ELMs) are plasma instabilities that occur in most high-confinement mode (H-mode) plasmas[1]. ELMs have the beneficial potential of flushing impurities from the core, however, they also can lead to significant power and particle fluxes to the machine's walls, particularly at the divertor plates, which can compromise their integrity[2], possibly leading to melting, erosion and recrystallization events[3][4]. The nominal projected values of peak energy fluence in ITER due to unmitigated type-I ELMs[5] was found to be above the W monoblock surface melting, posing a major operational constraint for ITER, SPARC and potentially any other future machines with W-based divertor tiles.

In the framework of future plasma scenarios, ELM-mitigation strategies are currently being studied and include, among others, operating in ELM-free regimes such as Quiescent H-mode[6], I-mode[7] or small/no ELM regimes[8]. A regime that is currently being considered for future operations involves high pedestal foot/separatrix collisionality plasma with ELMs that, compared to standard type-I ELMs, are characterized by lower energy losses, higher frequency and lower peak energy fluence near the strike points. This regime is consistent with the need for a power-exhaust solution in which (partially) detached divertors enable energy dissipation and maintain tolerable fluxes to the targets. These small ELMs, originally referred to as type-II ELMs or small ELMs, were first described in early 2000s works from JT-60U[9], ASDEX Upgrade (AUG)[10], DIII-D[11] and JET[12]. More recent works have shown how by increasing density and shape, these small ELMs become high-frequency filaments slightly above the background plasma while increasing upstream radial fluxes[13][14][15][16][17] and were subsequently re-named as the quasi-continuous-exhaust regime[18][19]. In this work, we will refer to small/type-II/QCE/filamentary ELMs interchangeably. Recent experimental[20] and modeling[21][19] studies from DIII-D and AUG found the small ELMs to originate from the pedestal foot. Projections based on those findings resulted in peak ELM energy fluence values that are within tolerable limits for the divertor target in ITER and SPARC[20]. However, a scaling for the ELM energy loss (relative to the plasma stored energy) for such a regime has yet to be proposed.

A pioneering ELM energy loss scaling for ITER-like scenarios was proposed by Loarte et al. [22], where it was shown that there is an inverse dependency between relative ELM energy losses normalized by the plasma stored energy and normalized neoclassical electron collisionality at the pedestal top, $\nu^*_e$. Such a study included mainly type-I ELMs, with some smaller type-III ELMs[23], in high-current low-q95 scenarios from JET, with some data points from DIII-D. More recent experimental work from DIII-D[24], based on pioneering studies such as[25], has shown that type-I ELMs in peeling-ballooning-limited discharges (at the 'nose' of the pedestal stability diagram) do not follow the relation reported by Loarte, indicating that $\nu^*_e$ might not be the most appropriate parameter to project ELM energy losses in future scenarios with peeling-ballooning-limited or ballooning-limited pedestals at high-edge-density (i.e. compatible with detachment).

In this study, we first evaluate the Loarte model over a multi-machine database of elevated q95 (> 3.5) type-I and small ELMs. That step is followed by regression analysis on ELM energy losses for type-I ELMs only. A comparison between experimental measurements and projected values via the ELM energy fluence model[5] is reported, together with results from pedestal stability analysis. Furthermore, the dependency between separatrix collisionality, ELM energy content and ELM frequency is discussed. Finally, a new scaling is proposed for small ELMs adopting normalized separatrix values and experimental data from DIII-D, AUG and TCV, and used to project to future devices.

*2. Methods*

A new analysis, involving a multi-machine database of plasma scenarios with a wide range of normalized ELM energy losses, ($\Delta E_{ELM}/W_{plasma}$) between 0.5% and 14%, is carried out and includes discharges spanning neoclassical electron collisionality at the pedestal top, $\nu^*_{e,neo}$[26], calculated as $\upsilon^*_{e,neo} = 6.92 \times 10^{-5} \frac{R_{major} q_{95} ln\Lambda n_e Z_{eff}}{T_e^2 \left(\frac{a_{minor}}{R_{major}}\right)^{3/2}}$ with $R_{major}$ [m] the major radius, $n_e$ [$10^{19}$ $m^{-3}$] the electron density at the pedestal top, $\ln\Lambda$ the Coulomb logarithm calculated as $ln\Lambda = 31.3 - \ln\left(\frac{\sqrt{n_e}}{T_e}\right)$ and found to be around 14-16 throughout the dataset, $a_{minor}$ [m] the plasma minor radius, $Z_{eff}$ the effective charge and Te [keV] the electron temperature at the pedestal top. $\nu^*_{e,neo}$ values in this study range from 0.1 to 10. Each $\Delta E_{ELM}/W_{plasma}$ data point is obtained by coherently averaging the signal over ~10 reproducible ELMs in steady-state scenarios (avoiding transient time windows as e.g. density ramps). The main plasma characteristics of the discharges adopted, such as pedestal top values of density and temperature, q95 and κ (plasma elongation), have been described in details in previous works for HL-2A[27], COMPASS[28], and with data from the JET-Carbon, JET-ITER-like-wall (ILW), MAST, and AUG (low $\nu^*_{e,neo}$) from Ref.[5]. New small ELM data on high-edge density elevated-q95 (q95 > 3.5) scenarios have been added, and are from AUG[16], DIII-D[20] and TCV[17][29]. For these latter cases, a detailed separatrix analysis is also included, while for the older scenarios, that is not available. As it will be shown later in the paper, this does not constitute a limitation to assess the results of the study, since large type-I ELMs are determined to originate and relate to parameters in a region near the pedestal top[5][24][30]. Note that AUG QCE cases are not available since the ELM energy loss could not be to date calculated with available diagnostics data. The scenarios examined in this study include both W- and C-wall machines, with minor radii from ~0.2 m to ~1.2 m and major radii from ~0.7 m to ~ 3 m. All the discharges adopted are lower single-null with ion B×∇B drifts toward the lower (main) divertor, and the toroidal magnetic field varies from 0.4 to 3.2 T.
The location of the separatrix is obtained by shifting the fitted Thomson Scattering pedestal profiles according to a corrected-Spitzer-Harm power balance[31] so that $T_e$ is obtained via $T_e = \left(\frac{7}{16} \frac{P_{sep} q_{cyl}^2 R_{maj}}{\kappa_0 \hat{\kappa} a \lambda_q}\right)^{2/7}$, where $P_{sep}$ is the power entering the SOL (calculated as $P_{sep} = P_{in} - P_{rad,core}$), $q_{cyl}^2$ the safety factor in the cylindrical approximation[32], $\hat{\kappa}$ the effective elongation[33], $\kappa_0$ the Spitzer-Harm conduction coefficient[31] and $\lambda_q$ the SOL heat flux width mapped at the outer mid-plane (OMP). It should be noted that for the DIII-D and TCV cases, $\lambda_q$ is extracted from infrared (IR) measurements of the outer divertor target (mapped at the OMP to account for flux expansion at the divertor), while in the AUG cases the relation $\lambda_q = 2/7\lambda_{T_e}$ is used [34].
IRTV divertor heat flux measurements at DIII-D[35] allow to resolve intra-ELM heat flux for both large type-I and small ELMs (integration time of 62 μs), from which the peak ELM energy fluence can be extracted via the relation $\varepsilon_{||,measured} = \frac{1}{\sin(a_{osp})}\int_{t=t_0}^{t=t_{ELM}}[q_{\perp,div}(t) - q_{\perp,div\ inter-ELM}(t)]dt$ , where $a_{osp}$ is the incident angle of the magnetic field at the divertor outer target surface (typically 1-3°) and $q_{\perp,div}$ the peak perpendicular heat flux at the outer divertor. This quantity, $\varepsilon_{||,measured}$, can be compared with the Eich model[5], which assumes the pedestal top is connected to the floor and being emptied out via parallel transport by the ELM, and it is written as $\varepsilon_{modelled} = 6\pi p_e a_{minor} B_{tor,omp}/B_{pol,omp}\sqrt{\frac{1+\kappa^2}{2}}$, where $p_e$ is the electron pressure at the pedestal top, $B_{tor,omp}$ and $B_{pol,omp}$ the toroidal and poloidal magnetic field at the OMP and κ is plasma elongation.

## *3. Results*

### *3.1 Assessing the applicability of the Loarte's scaling*

In Fig. 1 the ELM energy loss is shown as a function of $\nu^*_{e,\,neo}$ at the pedestal top in log-log space for the database examined, where the ELMs with $\Delta E_{ELM}/W_{plasma} > 2\%$ (labeled as large ELMs) and those with $\Delta E_{ELM}/W_{plasma} < 2\%$ (labeled as small ELMs) are separated by dashed boxes. This 2% threshold is an empirical operational definition which is justified by different ELM characteristics and pedestal stability scenarios, as it will be discussed throughout the manuscript. By inspecting the large (or type-I) ELMs alone, it can be seen that no clear dependency with $\nu^*_{e,\,neo}$ is apparent: a linear and a power law fit to the data (not shown) give coefficients of determination $R^2$=0.02 and $R^2$=0.05, respectively. This is in line with a previous studies at DIII-D[24], where the Loarte model was tested against peeling-ballooning-limited type-I ELM scenarios, and it was determined to be not applicable. By simply projecting the foreseen value for SPARC's $\nu^*_{e,\,neo}$ of ~0.11[36], a value of $\Delta E_{ELM}/W_{plasma}$ ~6±2% would result.
For the small ELM cases in Fig. 1 (highlighted in a dashed black square), although the scatter in the data is such that no definitive pattern can be quantified, the ELM losses are observed to somewhat decrease with increasing $\nu^*_{e,\,neo}$ for the available dataset. A more in-depth discussion on the small/filamentary ELMs and their relation with the pedestal and separatrix plasma parameters follows later in the paper.

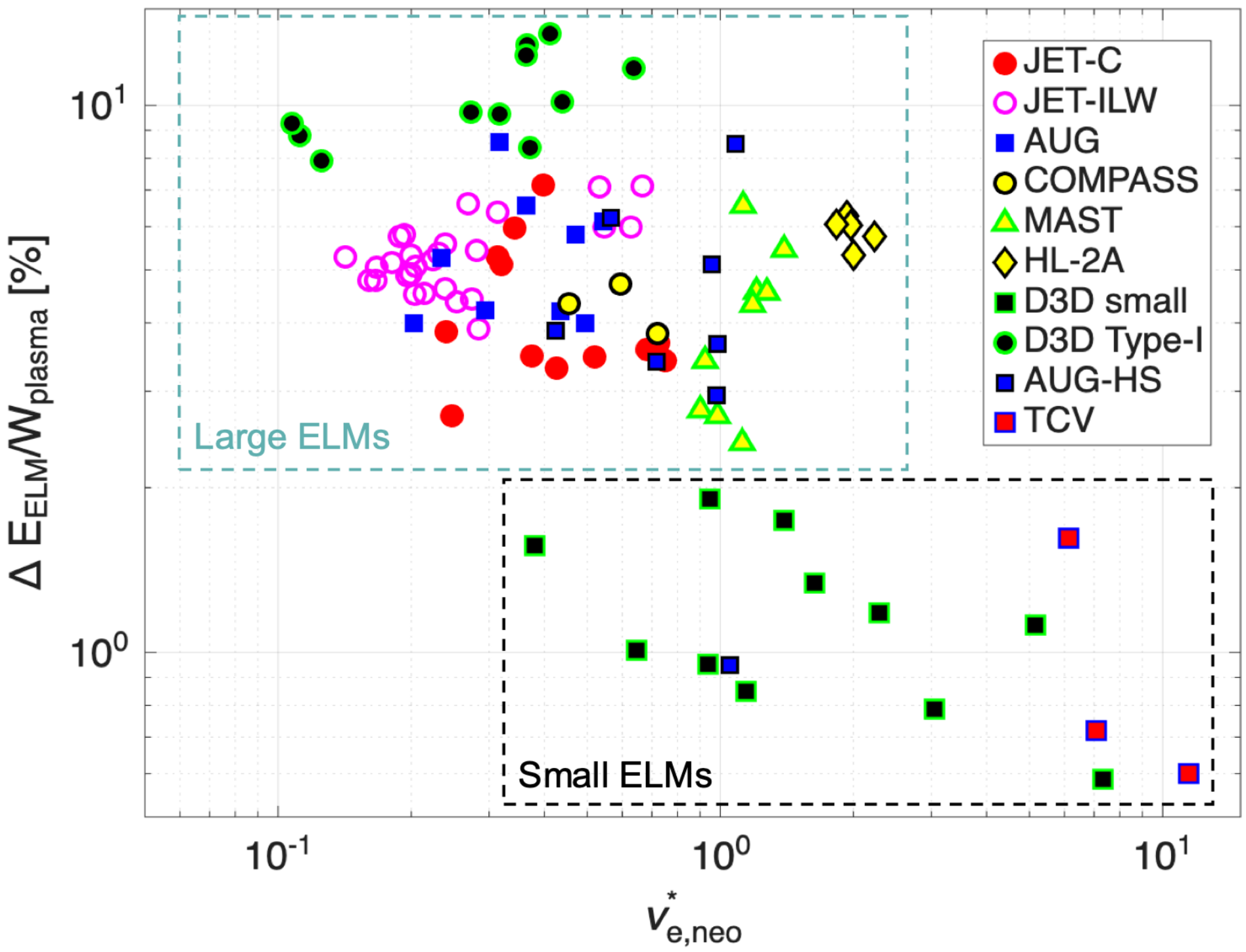

**Figure 1. Normalized ELM energy loss, $\Delta E_{ELM}/W_{plasma}$, as a function of the normalized electron collisionality at the pedestal top for various machines, plotted in log/log space. A distinction between large and small ELMs in their extents in each dimension is indicated by green and black dotted squares, with the small ELMs having $\Delta E_{ELM}/W_{plasma} < 2\%$. AUG data are distinguished between standard and highly-shaped (HS) scenarios.**

### *3.2 Regression analyses and type-I ELM losses extrapolation to SPARC and ITER*

An empirical scaling for the ELMs examined in this work is attempted via least-squares regression analysis. The parameters chosen for the regression are (i) $T_e$ and $n_e$ from the pedestal top (ii) the major radius ($R_{maj}$) to account for the machine size and (iii) elongation (κ). The latter is included as high κ shaping is found to stabilize the equilibrium against large ELM events[37] by enhancing magnetic-well (favorable curvature) and the Pfirsch-Schluter contribution[38]. Fig. 2 (left) reports the regression result applied to the whole ELM dataset as shown in Fig. 1, where a clear distinction on ELM classes based on ELM size is now observed: Moving further toward small ELMs leads to the data deviating more from the projected value. This results in an empirical scaling of the form $\left(\frac{\Delta E_{ELM}}{W_{plasma}}\right)[\%] = 5.6 * T_{e,ped}^{0.3} n_{e,ped}^{-0.6} \kappa^{-0.3} R_{major}^{0.05}$, with an overall low $R^2 = 0.2$. The lack of agreement informs on the presence of two distinct classes of ELMs, whose normalized ELM losses cannot be properly described with the same set of parameters. If the class of large ELMs with $\Delta E_{ELM}/W_{plasma} > 2\%$ is isolated and fit independent of the small ELMs, the resulting regression (in Fig. 2, right) now leads to a significantly better scaling ($R^2$ = 0.66,), where $\left(\frac{\Delta E_{ELM}}{W_{plasma}}\right)_{Type-I}[\%] = 6.8 *$ $T_{e,ped}^{0.03} n_{e,ped}^{-0.4} \kappa^{-0.4} R_{major}^{0.4}$. Other regressions including other key parameters i.e. $B_t$, q95 were attempted to improve the result, however yielding worse results than the one reported here. The regression reported suggests that the ELM energy loss is reduced with increasing pedestal density and plasma elongation, consistent with enhanced stabilization of the peeling drive due to increased edge collisionality, hence the $n_{e, ped}^{-0.4}$ dependency in the large ELM-only fit, and improved shaping stabilization of the peeling-ballooning boundary, hence the $\kappa^{-0.4}$ dependency. These factors can shift the plasma toward the ballooning-limited boundary, leading to high toroidal mode number (n) fast crash-recovery cycle, avoiding the strong pedestal collapse characteristic of type-I ELMs[39] (the pedestal stability analysis for this observation is reported in the next Section). The $T_{e,ped}$ sensitivity within the fitted dataset can be neglected (exponent of 0.03), while the positive $R_{maj}$ dependence reflects a combination of plasma size effects. Applying the large-ELM fit for a scenario of type-I ELMs in SPARC[36] with $T_{e, ped}$ = 3.9 keV, $n_{e, ped}$ = 3E20 $m^{-3}$, $\kappa$ = 1.97 and $R_{maj}$ = 1.85 m, a value of $\Delta E_{ELM}/W_{plasma}$ = 4.5% is obtained. For an ITER 15 MA Q~10 scenario[40] with $T_{e, ped}$ = 5 keV, $n_{e, ped}$ = 8.5E19 $m^{-3}$, $\kappa$ = 1.85 and $R_{maj}$ = 6.2 m, the projected $\Delta E_{ELM}/W_{plasma}$ is ~12%. Both values are likely intolerable for the integrity of the divertor targets. Note that the input parameters used for these extrapolations should be considered indicative, since future studies with more robust datasets on scenario development and pedestal profiles projections to future machines may provide different figures than the ones adopted here.

The simple exercise reported in this Section indicates the presence of two clearly distinct classes of ELMs. Regarding large type-I ELMs only, the empirical scaling extracted will be challenged with further scenarios (possibly from other machines as well) to evaluate and better quantify the accuracy of the extrapolations reported here.

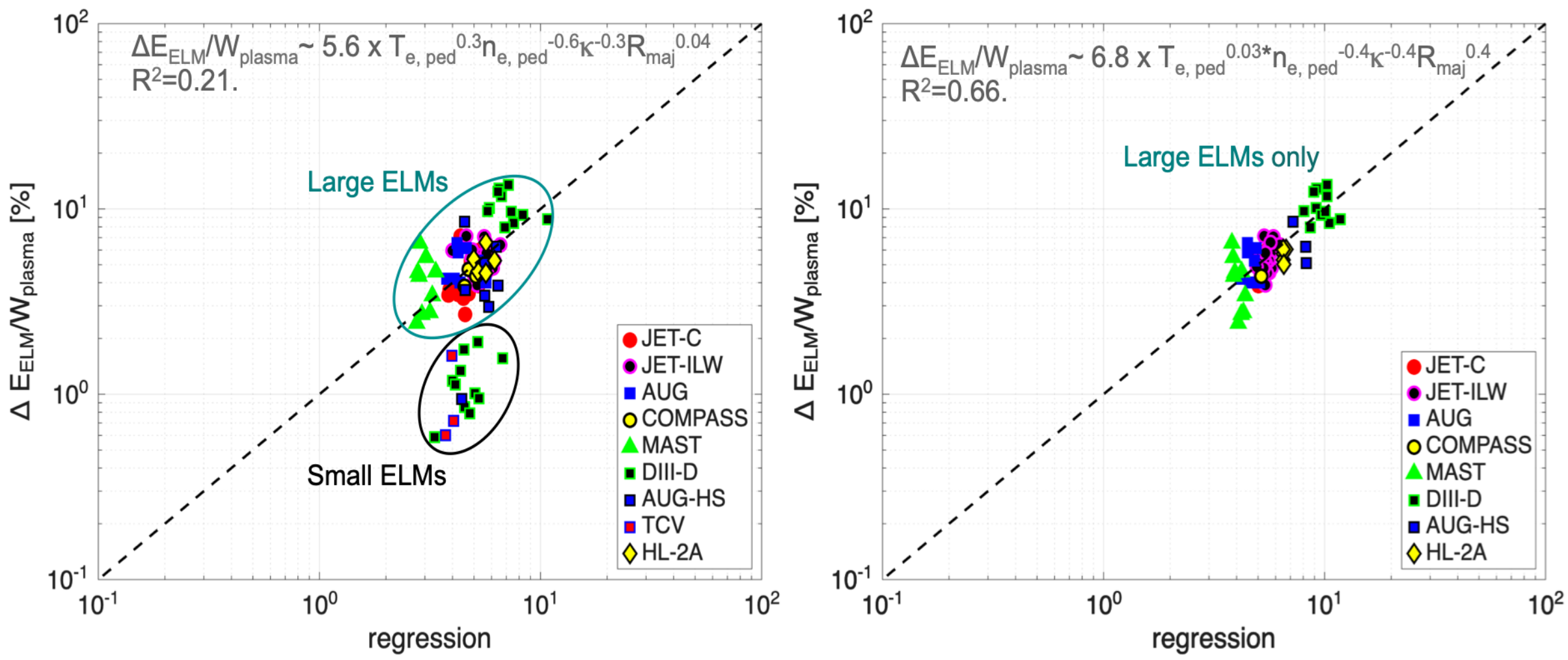

**Figure 2 Regression analysis result as applied to the full ELM database (left), and applied only to large-ELMs (right).**

*3.3 Energy fluence and pedestal stability analysis for different ELM classes*

Experimental studies from DIII-D and AUG[20] and simulations with the BOUT++[41] and HELENA[21] codes have inferred that small (or type-II) ELMs originate at the pedestal foot without affecting the electron pressure pedestal structure except for a region near (and across) the separatrix. This type of instability is due to unstable ballooning modes that are observed at high edge collisionality; when the separatrix density keeps increasing (e.g. via D2 fueling), the ELMs increase in frequency and decrease in amplitude and, beyond a certain threshold (which will be discussed later), the small/filamentary/QCE ELM scenario is accessed. Analysis on the peak parallel ELM energy fluence ($\varepsilon_{||}$) has shown that experimental $\varepsilon_{||}$ values of small ELMs can be reproduced by the Eich model[5] only if electron pressure values from the pedestal foot/separatrix are used[20]. For large type-I ELMs, pedestal top values of electron pressure were found to work best to reproduce the experimental data[24][27][28].
These observations suggest that the key pedestal region governing the main features that characterize the small ELMs in high-density scenarios lies in close proximity to the separatrix, and not the pedestal top. Within the framework of the *separatrix operational space*[42][43], or SepOS, edge plasma transport parameters are characterized at the separatrix via the turbulence control parameter $\alpha_t$. Defined as $\alpha_t = KRq_{cyl}^2 Z_{eff} n_{e,sep}/T_{e,sep}^2$, with $K \sim 3 * 10^{-18}$, $\alpha_t$ is an engineering parameter obtained from the global turbulence parameter in Scott[44]. This normalized parameter is representative of the relative importance of the resistive ballooning instability over the drift-wave one, and is derived from $\alpha_t = \omega_B C$ i.e. the product between the strength of the interchange turbulence due to magnetic curvature $\omega_B = \frac{2<\lambda_p>}{R}\left(1 + \frac{1}{\bar{Z}}\right)$, with $<\lambda_p>$ the poloidally-averaged separatrix pressure decay length and $\bar{Z}$ the average ion charge, and the normalized collisionality $C = 0.51\nu_{ei}\frac{<\lambda_p>}{c_s}\left(\frac{m_e}{M_i}\right)\left(\frac{\hat{q}_{cyl}R}{<\lambda_p>}\right)^2$, where $M_i$ is the effective ion mass and $\nu_{ei}$ the electron-ion collision frequency. Further details and explicit derivations can be found in [32].
The ratio between the experimental peak parallel energy fluence and that obtained via application of the Eich model for available DIII-D data is reported in Fig. 3a as a function of the separatrix turbulence parameter $\alpha_t$. Results from pedestal stability analysis with the code ELITE[39] are shown in Fig. 3b, where numbered cases correspond to the same number in Fig. 3a. To capture the effect of experimental and fitting uncertainties on the calculations of ELM stability with ELITE, we employ the new bouquet package within the open-source TokaMaker tool set[45] to produce an ensemble of kinetic equilibrium reconstructions representative of the experimental uncertainties. Self-consistent Grad-Shafranov solutions are produced within measured kinetic and toroidal current profile uncertainty envelopes and then passed into ELITE for calculation of the peeling-ballooning stability boundary. In this study, ELITE simulations are carried out for DIII-D cases only. In Fig. 3a it is shown that for values of $\alpha_t < \sim 0.3$ (denoted by a vertical black dashed line), the experimental $\varepsilon_{||}$ is reproduced by the model within ~50% (as shown by the grey region in Fig 3b.I). With low values of $\alpha_t$, ELITE simulations indicate a peeling-ballooning-limited boundary, as is typical for standard type-I ELMs, and a principal toroidal mode number n = 10 (Fig. 3b.I). When $\alpha_t > \sim 0.35$, the ELM energy fluence model over predicts the experimental values, indicating the presence of small ELMs. Such an overestimation increases monotonically with $\alpha_t$, reaching its minimum at $\alpha_t \sim 1$.

ELITE results inform that, once moving from type-I to small ELMs, the pedestal stability boundary moves down to the ballooning side, and the toroidal mode number n increases to n ~ 20-25 (Fig. 3b.II, 3b.III and 3b.IV).

Results reported here indicate a transition from type-I ELMs to small ELMs begins at $\alpha_t$~0.35, which is somewhat below, but in the same general range, with the $\alpha_t$~0.45 threshold for type-I to QCE regime in AUG, as shown in Fig. 6 in Ref. [18], where QCE scenarios are increasingly found for values above $\alpha_t$~0.3. This transition region should be accounted for when projecting the parallel peak energy fluence to future scenarios in high-separatrix-collisionality regimes.

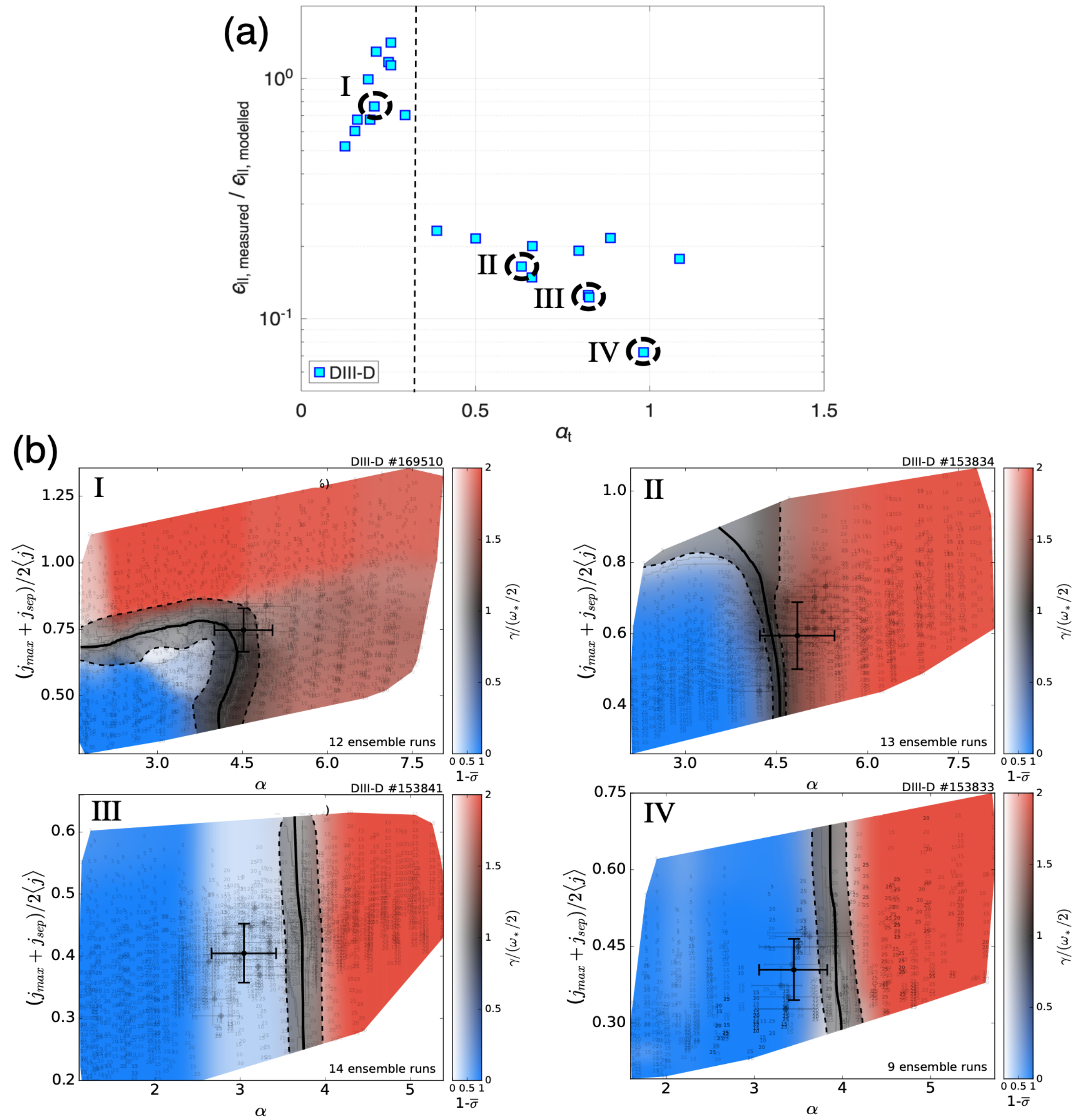

**Figure 3 (a) Ratio between the experimentally measured peak ELM energy fluence ($\varepsilon_{||,\,measured}$) and the one modeled with the peak ELM energy fluence model ($\varepsilon_{||,\,modelled}$) plotted versus the turbulence control parameter at the separatrix, $\alpha_t$. (b) Results from ELITE simulations, with the roman numbers indicating the respective case from Fig. 3a. The color bar represents the ELM instability growth rate normalized by the diamagnetic stabilization. Note that for each case, several runs are performed, as indicated by the number of ensemble runs in each plot. The thick black line is the mean of the stability boundaries across the ensemble of runs and the dashed black lines are the ±1 normalized standard-deviation boundaries of the ensemble stability limit.**

*3.4 Relating ELM frequency with ELM size and separatrix conditions*

To further characterize how ELM characteristics vary within the ELM classes examined, the relation between ELM frequency ($f_{ELM}$) with (i) $\alpha_t$ and (ii) ELM energy loss ($\Delta E_{ELM}/W_{plasma}$) is reported in Fig. 4a and Fig. 4b, respectively, for DIII-D, AUG and TCV cases. When inspecting Fig. 4a, two main regions can be identified, with $\alpha_t \sim 0.35$ marking the threshold for the transition: For $\alpha_t < \sim 0.35$ $f_{ELM}$ increases steeply, from ~10 Hz to ~100 Hz, while when $\alpha_t$ ranges between 0.4 and 1.35, $f_{ELM}$ rises slowly, from 200 Hz to 1 kHz. The small ELM data can be well fitted ($R^2$=0.83) with an exponential function, leading to the scaling:

$$f_{small_ELMs} = 46e^{(2.25*\alpha_t)},$$

shown with a black dotted line in Fig. 4(a). Fitting on the type-I ELM cases have been attempted, but resulted in low quality fits, indicating that the $\alpha_t$ parameter regulates the small ELM dynamics, as it will be discussed further in the next Section together with projections to SPARC and ITER.

In Fig. 4b it can be seen that the ELM frequency decreases slowly up to $\Delta E_{ELM}/W_{plasma}$~8%, before dropping sharply at $\Delta E_{ELM}/W_{plasma}$~10%. This finding is qualitatively similar to what has been reported for type-I ELMs in JET [22]. The resulting trend for the data analyzed here is consistent with the small/QCE ELMs having a fast cycle (reflected in the high $f_{ELM}$) that prevents a strong collapse of the pedestal (i.e. low $\Delta E_{ELM}/W_{plasma}$), unlike for the more common type-I ELMs, where a strong collapse of the pedestal structure is observed[25].

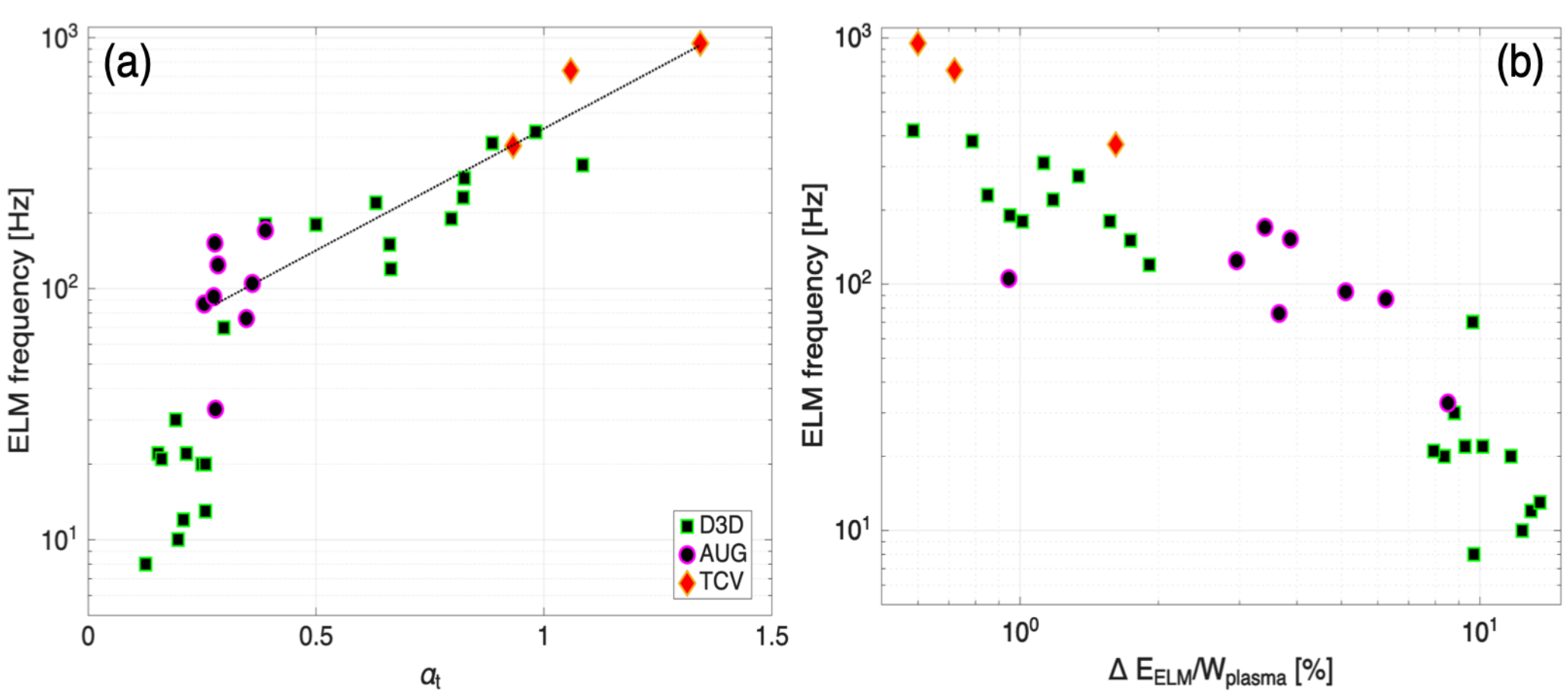

**Figure 4 (a) ELM frequency VS $\alpha_t$ (a) and $\Delta E_{ELM}/W_{plasma}$ (b) for the small ELM scenarios examined in this study. Type-I ELM cases for DIII-D and AUG are also included. Dotted linein Fig. 4(a) is the exponential fitting to the data.**

*3.5 Small ELM energy loss scaling and extrapolation to SPARC and ITER*

The normalized ELM energy losses for various ELM scenarios among DIII-D, AUG and TCV are reported in Fig. 5 as a function of the separatrix turbulence parameter $\alpha_t$. A clear distinction among two ELM classes can be observed: For the large ELMs, with $\alpha_t < \sim 0.3$, the energy losses are stable around $\Delta E_{ELM}/W_{plasma} \sim 10\%$ i.e. no dependence with separatrix parameters can be inferred. When transitioning to the small/filamentary ELM regime, a steep decrease in the ELM energy occurs between $\alpha_t > 0.25$ and $\alpha_t < 0.4$. In the dataset analyzed, only DIII-D data cover the whole $\alpha_t$ range, while the data from TCV and AUG fill key regions in the parameter space i.e. TCV data inform on high $\alpha_t$ conditions, whereas AUG data cover the narrow region of the transition. A similar narrow transition has been recently reported for BOUT++ simulations on high-collisionality small ELM regime at DIII-D[41]. From Fig. 5 it can be seen that once the small ELM regime is accessed, the ELM energy loss decreases with increasing $\alpha_t$ before single instabilities become impossible to resolve with an IRTV diagnostic, beyond $\alpha_t \sim 1$. This occurs in DIII-D and TCV cases, as highlighted within a blue square in Fig. 5: Since multiple ELMs (or filaments) are measured by filterscopes signals (lower-outer divertor D-$\alpha$) within the same stored energy drop, those data points are obtained by distributing the stored energy drop evenly among the number of ELM peaks that occur within that time window. Single ELMs are well resolved in all the other data points.
The dependence of the relative small ELM energy loss on $\alpha_t$ can be parameterized with an exponential decay function of the form $y = Ae^{-\left(\frac{x}{B}\right)}$, resulting in

$$\left(\frac{\Delta E_{ELM}}{W_{plasma}}\right)_{Small} [\%] = 1.6e^{-\left(\frac{\alpha_t}{2}\right)},$$

where the decay scale $B$=2 constitutes a relatively gradual decrease if compared to the much steeper decay ($B$=0.13, not shown in Fig. 5) that characterizes the data in the region transitioning from large to small ELMs (indicated in grey in Fig. 5).
For a SPARC scenario in the QCE regime[43] with $n_{e,sep}$=4E20m$^{-3}$, $T_{e,sep}$=156 eV, $q_{cyl}$=3, $Z_{eff}$=1 and $R_{major}$=1.85m, a value of $\alpha_t = 0.86$ is obtained, which corresponds to a relative ELM energy loss of 1.04% (indicated in yellow in Fig. 5), and, using the ELM frequency scaling reported above, a $f_{ELM}$~320 Hz. If $\alpha_t$ is pushed beyond 1, $\Delta E_{ELM}/W_{plasma}$ falls below 1% and $f_{ELM}$ goes above 430 Hz (reaching 1 kHz at $\alpha_t$ =1.38).
For the ITER D-T 15 MA scenario at $f_{GW}$=0.5 (Table 22, column A in[46]) (run ID: 135011-7), with $n_{e,sep}$=4.9E19m$^{-3}$, $T_{e,sep}$=200 eV [47], $q_{cyl}$=3.1, $Z_{eff}$=1 and $R_{major}$=6.2 m, the resulting $\alpha_t = 0.23$ falls below the transition threshold ($\alpha_t$>~0.35) between large and small ELMs observed in the dataset. ITER high-fuelling (ITER high fuel. in Fig. 5) scenarios examined in [46] report $n_{e,sep}$ values up to 8.3E19m$^{-3}$ (Table 22 column C in[46]), which, assuming similar input as the baseline case, would lead to $\alpha_t = 0.64$ i.e. well into the small ELM regime, leading to $\Delta E_{ELM}/W_{plasma} \sim 1.2\%$ (indicated by a cyan cross in Fig. 5) and $f_{ELM}$~200 Hz. It should be also noted that a recent study [48] also supports the applicability of the small ELMs/QCE regime in ITER and EU-DEMO, if sufficiently high separatrix density is achieved in each ($n_{e,sep}/n_{gw}$~ 0.3).
In the extrapolations to future machines reported here, an important caveat should be taken into account: Although $Z_{eff}$=1 has been assumed, reactor-relevant operation will likely require extrinsic impurity seeding to radiate a significant fraction of the power entering the SOL. The resulting increase in the separatrix $Z_{eff}$ would raise $\alpha_t$, relative to the values reported here.

The projections reported in this Section should be regarded as a first attempt to provide quantitative estimation of ELM energy losses in reactor-relevant conditions. Future analysis on broader datasets from current machines will be necessary to improve the accuracy of these extrapolations for future fusion scenarios.

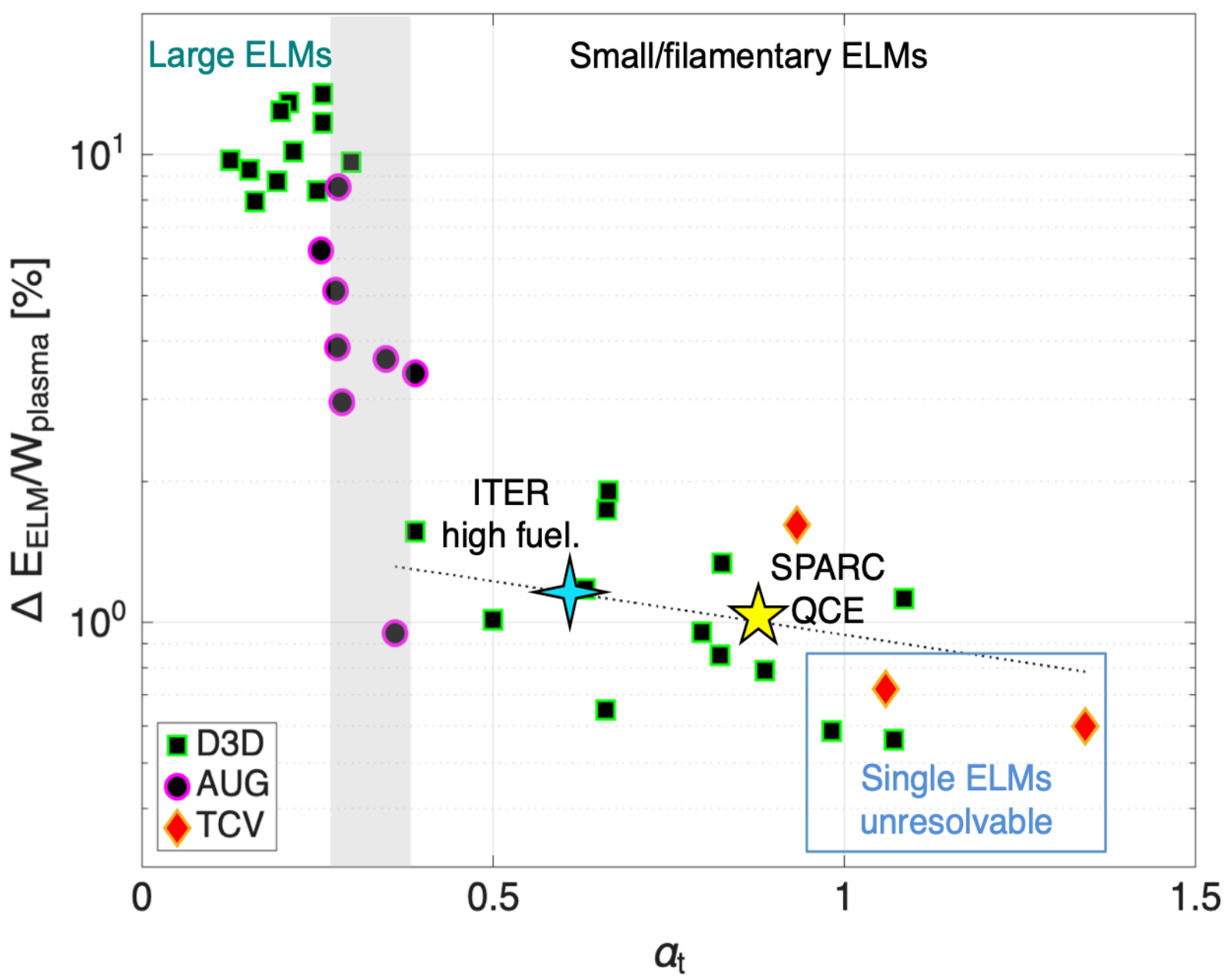

**Figure 5** **$\Delta E_{ELM}/W_{plasma}$ VS $\alpha_t$ for the available DIII-D, AUG and TCV data. The transition between large and small/filamentary ELM regimes is indicated in shaded grey area, and the projected value for SPARC-QCE is reported in yellow (star) and ITER high-fuelling case in cyan (cross). The exponential decay fitting to the small ELM data is indicated with a dotted line. Note that no dependency between $\Delta E_{ELM}/W_{plasma}$ and $\alpha_t$ is found for large ELMs.**

## *4. Summary and Conclusions*

The main findings reported in this work can be summarized as follows:

- A multi machine database including large type-I and small ELM regimes has been compiled to assess the applicability of existing ELM energy scaling (Loarte scaling).
- The pedestal-top neoclassical collisionality does not provide a robust ordering parameter for $\Delta E_{ELM}/W_{plasma}$ when elevated-q95, peeling-ballooning-limited type-I ELMs and ballooning-limited small ELMs are considered.
- A regression analysis for the relative ELM energy losses of type-I ELMs-only has been presented. The ELM losses are found to decrease with increasing pedestal-top density and plasma shape. When projecting $\Delta E_{ELM}/W_{plasma}$ for nominal scenarios at SPARC and ITER (15 MA Q=10 D-T), values of 4.5% and 12% are obtained, respectively, which lends to concerns about survivability of plasma-facing materials with frequent type-I ELM events.

- A comparison between the measured peak parallel ELM energy fluence and the Eich model indicates that large type-I ELMs are consistent with plasma losses connected to the pedestal-top region and peeling-ballooning stability limits.
- However, small ELMs are found to be governed primarily by plasma conditions near/at the separatrix, consistent with filaments originating at the pedestal foot (and not the pedestal-top), indicating the separatrix region to be more suitable to extrapolate ELM scenarios.
- The separatrix turbulence parameter, $\alpha_t$, is shown to order the transition from large type-I ELMs to small ELMs. As $\alpha_t$ increases, the pedestal moves toward a ballooning-limited regime, the toroidal mode number increases, the ELM frequency rises via the relation $f_{small_ELM} = 46e^{(2.25*\alpha_t)}$, and the normalized ELM energy loss decreases.
- A first attempt on an empirical scaling for small-ELMs-only has been presented. For SPARC QCE-relevant values of $\alpha_t > 0.7$[43], a value of $\Delta E_{ELM}/W_{plasma} \sim 1\%$ is obtained, while for ITER high-fuelling case, $\Delta E_{ELM}/W_{plasma} \sim 1.2\%$. It is shown that if $\alpha_t > 0.9$ is achieved, the ELM energy loss is projected to fall below 1%. These figures are considerably more favorable for plasma-facing survivability in a small ELMing scenario compared to a Type-I ELMing scenario.

In this letter, two distinct ELM classes are identified and treated accordingly to provide extrapolations to SPARC and ITER scenarios. Large/type-I ELM losses are found to scale with plasma parameters at the pedestal top and plasma shape, while small/filamentary/QCE ELMs with separatrix conditions. This picture is consistent with the different pedestal origin of these instabilities and motivates the development of two separate scaling laws. In particular, high-density small ELMs energy losses should be projected to future scenarios using separatrix-based quantities, with $\alpha_t$ providing a more suitable knob to control ELM energy losses than the pedestal top collisionality. Future work will extend the current database and will include dedicated separatrix analysis from other machines, such as JET, Alcator C-MOD, MAST-U, aiming to further test and improve the relative ELM energy loss scalings reported here.

*5. Acknowledgments*

This material is based upon work supported by the U.S. Department of Energy, Office of Science, Office of Fusion Energy Sciences, using the DIII-D National Fusion Facility, a DOE Office of Science user facility, under Award(s) DE-FC02-04ER54698, DE-SC0026433, DE-SC0026404, DE-SC0022270, DE-FG02-08ER54999 and DE-AC52-07NA27344.
This work has been carried out within the framework of the EUROfusion Consortium, partially funded by the European Union via the Euratom Research and Training Programme (Grant Agreement No 101052200 — EUROfusion). The Swiss contribution to this work has been funded by the Swiss State Secretariat for Education, Research and Innovation (SERI). Views and opinions expressed are however those of the author(s) only and do not necessarily reflect those of the European Union, the European Commission or SERI. Neither the European Union nor the European Commission nor SERI can be held responsible for them.
This work was supported in part by the Swiss National Science Foundation.

*6. Disclaimer*

*7. Data availability statement*

*8. Bibliography*

[1] F. Wagner et al., Phys. Rev. Lett. 49 (1982) 1408
[2] R. Pitts et al., Nucl. Mater. Energy 20 (2019) 100696
[3] J. Gunn et al., Nucl. Fusion 57 (2017) 046025
[4] J. Linke *et al* 2011 *Nucl. Fusion* **51** 073017
[5] T. Eich et al., Nucl. Mater. Energy 12 (2017) 84-90
[6] D. Ernst et al., Phys. Rev. Lett. 132 (2024) 235102
[7] A.E. Hubbard *et al* 2017 *Nucl. Fusion* **57** 126039
[8] E. Viezzer et al., Nucl. Mater. Energy 34 (2023) 1013082023
[9] Y. Kamada et al., Plasma Phys. Control. Fusion 42 (2000) A247
[10] J. Stober et al., Nucl. Fusion 41 (2001) 1123
[11] A. Leonard et al., J. Nucl. Mater 290-293 (2001) 1097-1101
[12] G. Saibene et al., Nucl. Fusion 25 (2005) 297
[13] E Wolfrum *et al* 2011 *Plasma Phys. Control. Fusion* **53** 085026
[14] B. Labit et al., Nucl. Fusion 59 (2019) 086020
[15] R. Perillo et al. 2025 Phys. Plasmas 32, 022501
[16] A. Redl et al. 2024 Nucl. Fusion 64, 086064
[17] A. Stagni et al. 2024 Nucl. Fusion 64, 026016
[18] M. Faitsch *et al.* 2023 *Nucl. Fusion* **63** 076013
[19] G.F. Harrer et al. 2022 Phys. Rev. Lett. 129, 165001.
[20] R. Perillo et al., J. Plasma Phys. 92 (2026) E35
[21] L. Radovanovic *et al* 2022 *Nucl. Fusion* **62** 086004
[22] A Loarte *et al* 2003 *Plasma Phys. Control. Fusion* **45** 1549
[23] Zohm H 1996 Plasma Phys. Control. Fusion 38 1213
[24] *Knolker* et al *2018* Nucl. Fusion 58 096023
[25] A. Leonard et al., Plasma Phys. Control. Fusion 44 (2002) 9

[26] O. Sauter et al., *Phys. Plasmas* 6, 2834–2839 (1999)
[27] J.M. Gao et al 2021 Nucl. Fusion 61 066024
[28] J. Adamek *et al* 2023 *Nucl. Fusion* **63** 086009
[29] A. Stagni et al. 2022 Nucl. Fusion 62, 096031.
[30] W. Fundamenski et al., Plasma Phys. Controlled Fusion 48 (2006) 109
[31] P. Stangeby et al., Nucl. Fusion 55, 093014
[32] T. Eich et al., Nucl. Fusion 60 (2020) 056016
[33] N.A. Uckan et al., Proceedings from The 14th IEEE/NPSS Symposium Fusion Engineering, 30 Sept. 1991 - 3 Oct. 1991, San Diego, CA, USA, id.146
[34] M. Faitsch et al., Plasma Phys. Control. Fusion 57 (2015) 075005
[35] C.J. Lasnier et al. 1998 Nucl. Fusion 38, 1225
[36] Hughes JW et al., *Journal of Plasma Physics*. 2020;86(5):865860504.
[37] J.M. Greene et al., General Atomics report, 1998. Link: https://fusion.gat.com/pubs-ext/ComPlasmaPhys/A22135.pdf
[38] J.W. Connor et al., Phys. Plasmas, Vol. 5, No. 7, July 1998
[39] P. B. Snyder et al., Nucl. Fusion 44 (2004) 320–328
[40] Luda et al. Nucl. Fusion (2025) 072001
[41] N. Li Nucl. Fusion 65 076023
[42] T. Eich *et al* 2021 *Nucl. Fusion* **61** 086017
[43] T. Eich et al., Nucl. Mater. Energy 42 (2025) 101896
[44] B. Scott et al., *Phys. Plasmas* 12, 062314 (2005)
[45] https://zenodo.org/records/19411820
[46] Garzotti et al., Nucl. Fusion 59 (2019) 026006
[47] X.X. He et al 2022 Nucl. Fusion 62 056003
[48] M. Dunne *et al* 2024 *Nucl. Fusion* **64** 124003